\documentclass[a4paper,fleqn,usenatbib]{mnras}

\usepackage[T1]{fontenc}
\usepackage{ae,aecompl}

\usepackage{graphicx}	
\usepackage{amsmath}	
\usepackage{amssymb}	

\title[K band LFs in Clusters]{The $K$-band luminosity functions of cluster galaxies}

\author[R. De Propris]{
Roberto De Propris$^{1}$\thanks{E-mail: rodepr@utu.fi}
\\
$^{1}$FINCA, University of Turku, V{\"a}is{\"a}l{\"a}ntie 20, Piikki{\"o}, 21500, Finland\\
}

\date{Accepted XXX. Received YYY; in original form ZZZ}

\pubyear{2016}

\begin{document}
\label{firstpage}
\pagerange{\pageref{firstpage}--\pageref{lastpage}}
\maketitle

\begin{abstract}

We derive the galaxy luminosity function in the $K_s$ band for galaxies in 24 clusters to provide
a local reference for higher redshift studies and to analyse how and if the luminosity function varies
according to environment and cluster properties. We use new, deep $K$ band imaging and match 
the photometry to available redshift information and to optical photometry from the SDSS or the
UKST/POSS: $>80\%$ of the galaxies to $K \sim 14.5$ have measured redshifts. We derive
composite luminosity functions, for the entire sample and for cluster subsamples . We consider the
luminosity functions for red sequence and blue cloud galaxies. The full composite luminosity function 
has $K^*=12.79 \pm 0.14$ ($M_K=-24.81$) and $\alpha=-1.41 \pm 0.10$. We find that $K^*$ is 
largely unaffected by the environment but that the slope $\alpha$ increases towards lower mass clusters 
and clusters with Bautz-Morgan type $<$ II. The red 
sequence luminosity function seems to be approximately universal (within errors) in all environments: 
it has parameters $K^*=13.16 \pm 0.15$ ($M_K=-24.44$) and $\alpha=-1.00 \pm 0.12$ (for all
galaxies). Blue galaxies do not show a good fit to a Schechter function, but the best values for its
parameters are $K^*=13.51 \pm 0.41$ ($M_K=-24.09$) and $\alpha=-1.60  \pm 0.29$: we do not
have enough statistics to consider environmental variations for these galaxies. We 
find some evidence that $K^*$ in clusters is brighter than in the field and $\alpha$ is steeper, but 
note this comparison is based (for the field) on 2MASS photometry, while our data are 
considerably deeper.
\end{abstract}




\section{Introduction}
The galaxy luminosity function (hereafter LF) provides a fundamental description of the gross properties
of galaxy populations. The first task of theories of galaxy formation and evolution is to match the observed
LF, a task that has been somewhat difficult as uncertain baryonic effects (e.g., star formation) and feedback 
are needed to transform the mass function of dark matter halos into observables (e.g., see \citealt{Contreras2016, Narayanan2016} and references therein).

Clusters of galaxies, as the largest virialised systems in the Universe, have played an important role in
this field. The LF of cluster galaxies can be determined, even at high redshifts, via simple photometry, as
the overdensity with respect to the surrounding fields allows us to correct for contamination (from non
members in the foreground and background) by statistical means, without expensive redshift surveys. 
However, the faint end of the LF is still in dispute, even at low redshifts, as the steeply rising field counts
lead to progressively more unfavourable statistics. Studies of nearby clusters have claimed that the 
LF consists of a Schecher function at the bright end and a steeply rising power law at the faint end (e.g.,
\citealt{Moretti2015,Lan2016}) but others have found a single Schechter function (e.g., \citealt{Rines2008,
Sanchez2016}).

Ideally, the LF should be measured in a band where luminosity matches stellar mass as closely as possible,
in order to better compare with the predictions of theoretical models and avoid the effects of star formation
on bluer (optical) bands. The infrared $K$ band has been shown to provide a reasonable approximation to
the underlying stellar mass function, and even dynamical mass \citep{Gavazzi1996,Bell2001}. In addition,
both evolutionary and $k$-corrections are known to be small and to vary slowly with redshift or with galaxy
type. For these reasons, the $K$-band LF has been used as a probe of the evolution of galaxy populations
(e.g., see \citealt{Capozzi2012} and references therein).

There are relatively few local cluster LFs in the $K$-band, owing to the comparatively small size of 
infrared detectors until recently. In our previous work, we studied the Coma cluster using a complete
spectroscopic sample for its inner $25'$ \citep{DePropris1998}. \cite{Skelton2009} determined a LF for
the Norma cluster and \cite{Merluzzi2010} derived a composite LF for galaxies in the Shapley 
supercluster. Previously, \cite{DePropris2009} presented a composite LF for 10 clusters from the 
2dF sample of \cite{DePropris2003}. Here we determine the LF for 24 of these clusters, with new 
infrared imaging and high redshift completeness. The following sections describe the data and 
analysis, present the results and discuss these in the context of previous work and models for 
galaxy formation and evolution. Here, we assume the standard cosmological parameters
$\Omega_M=0.27$, $\Omega_{\Lambda}=0.73$ and $H_0=73$ km s$^{-1}$ Mpc$^{-1}$.

\section{Data and Analysis}

We have carried out deep $K$-band imaging for a set of 24 clusters from the sample of \cite{DePropris2003}
in order to derive composite $K$ band LFs. Our data consist of 300s images in the $K_s$ filter obtained 
at the CTIO 4m telescope, with either the Infrared SidePort Imager (ISPI -- \citealt{Probst2003}) or the 
NOAO Extremely Wide Field Infrared Mosaic  (NEWFIRM -- \citealt{Autry2003}) for 20 clusters, covering the 
clusters out to their Abell radius (1.5 $h^{-1}$ Mpc). For a few clusters (4/24) we have instead used 
available UKIDSS data from the Large Area Survey (Data Release 10) as we could not observe them from 
CTIO in the available time. Table~\ref{data} summarizes the data used and basic properties of the clusters. 

We have elected not to use 2MASS \citep{Skrutskie2006} photometry, except for purposes of calibration, 
as this is known to miss a considerable fraction of the flux for bright galaxies and to be incomplete for fainter 
ones \citep{Andreon2002,Kirby2008}. We confirm this by comparison with our photometry: on average, 2MASS
magnitudes (we use the homogeneous $14''$ aperture for reference, which should be large enough to include
all the flux) are systematically $\sim 0.3$ mag. fainter than ours. Further, some galaxies are already missing
from 2MASS (or misclassified as stars) at $K > 13$. However, our much deeper  data (300s on a 
4m class telescope, compared to the 52s exposures on a 1.3m telescope for 2MASS) should not
suffer from these issues. 

For ISPI and NEWFIRM imaging we observed using a five point dithering pattern. For NEWFIRM, the dithering
steps were large enough to remove the $\sim 30''$ gaps between the four detectors. Where the Abell radius was larger than the size of the detector, we mosaicked to cover the entire field (this took several ISPI fields, 
as the field of view is only $\sim 10'$, but only small NEWFIRM mosaics). ISPI data were reduced following 
the conventional pattern for infrared data: removal of flatfield with on and off dome light flats, median 
sky removal from neighbouring (in time) images and astrometric/distortion correction (from 2MASS 
stars in the field of view), followed by a median sum of the images. NEWFIRM uses a dedicated pipeline 
on specialised hardware; this is described in \cite{Swaters2009} and essentially carries out the infrared 
data reduction procedures in an automated fashion. The pipeline products are then placed in the NOAO 
archive for retrieval. Photometric calibration was carried out from 2MASS stars in each field.

For clusters within the UKIDSS sample, we used their photometry (Petrosian magnitudes) and star-galaxy 
classification. For ISPI and NEWFIRM data we carried out photometry with Sextractor \citep{Bertin1996} 
using a series of parameters that were found to be appropriate for galaxies in our previous work,  using
Kron-style magnitudes as returned by the software (MAG$\_$AUTO). We inspected visually all detections 
to remove contaminants (stellar spikes, trails, bad pixels on the detector edges, etc.) and confirm that
the catalog does not miss obvious sources or fragments bright ones. Star-galaxy separation was based 
on the Sextractor stellarity index, but we also confirmed the nature of all sources with reference to 
SDSS \citep{York2000,Eisenstein2011,Alam2015} or UKST/POSS (photographic) imaging. However,
we may miss compact dwarfs resembling M32, that have now been identified in significant numbers
in the CLASH sample \citep{zhang2016}, but may be misclassified as stars by lower resolution imaging;
these may affect the slope of the luminosity function.

UKIDSS data are in good agreement with our photometry at $K > 12$ -- the mean difference is a few 
hundreds of a  magnitude, which may be due to slight differences in the filter bandpasses. However, 
for galaxies at $K < 12$ in UKIDSS there is some evidence of missing flux (at the level of $\sim 0.2$ 
mag.) compared to our photometry, likely from the low surface brightness envelopes of brighter galaxies: 
even if UKIDSS is carried out on a 4m telescope in good conditions, the exposure times are necessarily
shallower than our deep pointed observations. 

Finally, we also obtained $g-r$ colours (using the Model magnitudes) for galaxies within the SDSS 
footprint and $B_J-R_F$ colours (as provided by the WFAU SSA service) for those with UKST data. 
See Table~\ref{data} for details of each source. Redshifts for all our galaxies were then retrieved from 
the NED database, with a $3''$ matching radius. The majority of redshifts come from the SDSS and the 
2dF \citep{Colless2001} surveys, but there are significant contributions from several other sources as well. 
Members were identified using the `double gapping' method originally proposed by \cite{Zabludoff1990}
and applied to our sample in \cite{DePropris2002}: galaxies were sorted in velocity space and the initial
sample of cluster members isolated from the field, requiring that the next nearest galaxy be at $cz > 1000$
km $s^{-1}$ (i.e., a first gap in the velocity distribution). We then computed a velocity dispersion and 
excluded galaxies separated by more than $1\sigma$ (second gap) to isolate a kinematically cleaned
sample of cluster members. From these we then compute the mean radial velocity and velocity dispersion
shown in Table~\ref{data}: the values we report are in good agreement with those presented in 
\cite{DePropris2003}. 

\begin{table*}
\caption{Sample of Clusters and properties}
\centering
\begin{tabular}{ccccccc}
\hline\hline
Cluster & RA (2000) & Dec (2000) & $cz$ & $\sigma$ & $K_s$ Source & Optical Source\\
            &    [hms]      &      [deg]     &  km/s &   km/s      & \\
\hline\\
Abell 930 & 10:06:46.27  & $-6$:11:18.0 & 17293 & 1033 & ISPI & UKST \\
Abell 954 & 10:13:44.89  & $-0$:07:13.2 & 28312 &  830  & ISPI & SDSS\\
Abell 957 & 10:13:38.28  & $-0$:55:31.5 & 13499 &  718  & NEWFIRM & SDSS\\
Abell 1139 & 10:59:17.80   & +1:09:13.0 & 11711 & 463 & UKIDSS & SDSS \\
Abell 1189 & 11:10:12.03 & +1:13:27.8 & 28780 & 786 & ISPI & SDSS\\
Abell 1236 & 11:22:44.9  & +0:27:44.0 & 30563 & 550 & UKIDSS & SDSS \\
Abell 1238 & 11:22:54.3  & +1:06:52.0 & 22145 & 573 & UKIDSS & SDSS\\
Abell 1364 & 11:44:28.56 & $-1$:50:07.6 & 32058 & 469 & ISPI & SDSS\\
Abell 1620 & 12:50:03.88 & $ -1$:32:25.0 & 25644 & 1042 & ISPI &SDSS \\
Abell 1663 &13:03:30.7 & $-2$:14:00.0 & 24921 & 751 & UKIDSS & SDSS\\
Abell 1692 & 13:11:36.8 & $-0$:28:59.0 & 22526 & 1073 & NEWFIRM & SDSS \\
Abell 1750 & 13:31:11.07 & $-1$:43:38.9 &  25484 & 1051 & ISPI & SDSS\\
Abell 2660 & 23:47:25.44 & $-25$:11:55.69 & 15919 & 719 & NEWFIRM & UKST \\
Abell 2734 & 0:11:21.63 & $-28$:51:15.55 & 18318 & 914 & NEWFIRM & UKST \\
Abell 2780 & 0:30:13.51 & $-29$:36:53.3 & 29783 & 990 & ISPI & UKST \\
Abell 3094 & 3:11:25.01 &  $-26$:55:52.20 & 20355 & 804 & NEWFIRM & UKST \\
Abell 3880 & 22:27:54.39 & $-30$:34:32.8 & 17322 & 733 & ISPI & UKST\\
Abell 4013 & 23:31:50.88 & $ -34$:03:19.95 & 16450 & 757 & NEWFIRM & UKST \\
Abell 4053 & 23:52:44.40 & $-28$:34:14.01 & 20195 & 1656 & NEWFIRM & UKST \\
EDCC 119 & 22:16:20.64 & $-25$:40:11.9 & 25400 & 1015 & ISPI & UKST \\
Abell S0003 & 0:03:11.13 & $ -2$7:52:42.41& 18984 & 939 & NEWFIRM & UKST \\
Abell S0084 & 0:49:22.83  &$ -2$9:31:12.1 & 32866 & 905 & ISPI & UKST \\
Abell S0166 &  1:34:14.70 &  $-3$1:38:56.09 & 20888 & 451 & NEWFIRM & UKST \\
Abell S1043 & 22:33:38.52 &  $ -2$4:45:50.97 & 11143 & 1449 & NEWFIRM & UKST \\
\hline\\
\end{tabular}
\label{data}
\end{table*}

\section{Luminosity Functions}

The individual LFs for each cluster are relatively poorly determined, because of small number
statistics (we have typically 60 members per cluster), especially at the bright end, even though
our redshift completeness (see below) is high. For this reason we produce a composite LF, 
following the methods outlined in \cite{Colless1989} and \cite{DePropris2003} and summarised
below. As this represents the average of several clusters spanning a wide range of properties,
it is likely to be a better measure of the LF than those derived for single clusters, but we explore
the variation of the LF according to cluster properties and for red and blue galaxies as well, to
understand the role of environmental variations.

As in our previous work we derive a LF at the mean redshift of the sample $z=0.075$. The reason 
for doing this is that in this way we avoid the uncertainty of carrying out $e$ and $k$ corrections to
$z=0$, which are somewhat poorly understood in the infrared (even though they are likely to be 
small, of the order of a few 1/100 of a mag.) and which of course would vary from galaxy to galaxy.
As the redshift difference between our clusters and $z=0.075$ is small, we can omit these corrections
as these are expected to be quite small, in a differential sense. 

Our procedure is as follows: we count galaxies in 0.5 mag. bins at $z=0.075$. For each cluster we 
calculate the difference in distance modulus between its redshift and $z=0.075$. We then count cluster 
members in apparent magnitude bins corresponding to the fixed magnitude bins at $z=0.075$. For instance, 
in Abell 1139 ($cz=11711$ km s$^{-1}$) the magnitude interval $10.47 < K < 10.97$ contributes to 
the galaxy counts in the $12.0 < K < 12.5$ bin at $z=0.075$, whereas in Abell S0084 ($cz=32866$ km
 $s^{-1}$) counts in the apparent magnitude bin $12.8 < K < 13.3$ contribute to the equivalent $12.0 
< K < 12.5$ counts at $z=0.075$. For the conventional cosmology, the distance modulus to this redshift 
is 37.60 mag. A similar approach is used for the SDSS LFs of \cite{Blanton2003} that are measured at 
$z=0.1$ and the red sequence LFs of clusters in \cite{Rudnick2009} where the reference redshift is 
$z=0.06$.

In order to create a composite LF we need to correct for incompleteness. Fig.~\ref{compl} shows the 
completeness fractions as a function of observed $K$ magnitude for all our clusters, including members, 
non members and objects for which no redshift is known. In general our spectroscopic completeness is 
well above 80\%, to at least $K \sim 14$.  We then correct for incompleteness and produce a composite 
luminosity function in the same manner as in \cite{DePropris2003}. In each magnitude bin, given $N_C$ 
as the number of spectroscopically confirmed cluster members, $N_R$ the number of galaxies with redshifts 
and $N_I$ as the total number of objects (including objects with no redshift) we find that the number of 
galaxies in magnitude bin $j$ of cluster $i$ is given by:

\begin{figure*}
  \includegraphics[width=\textwidth]{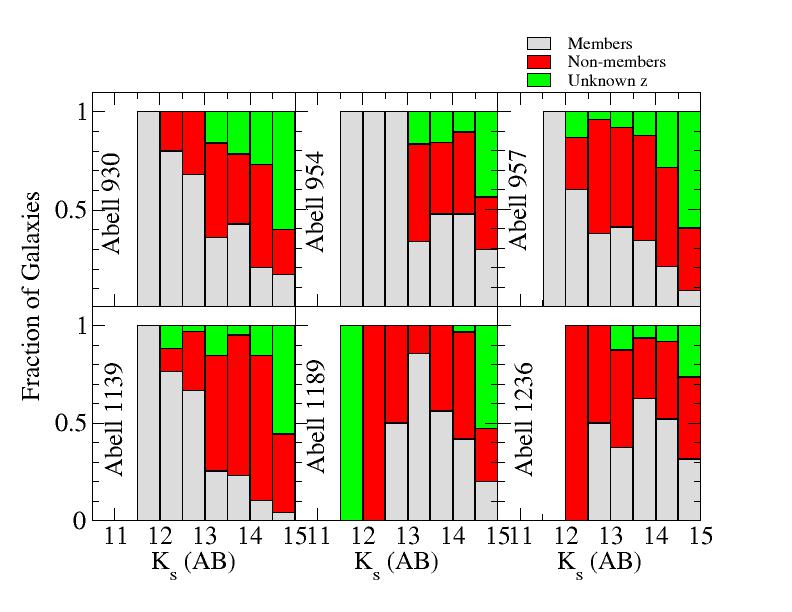}
  \caption{Redshift completeness histograms for a subset of the clusters in our sample (identified 
in the figure). We show members in  grey, non-members in red and objects with unknown redshift in green. 
See Appendix (online) for all the other clusters in the sample.}
\label{compl}
\end{figure*}

\begin{equation}
N_{ij}={N_C N_I \over N_R}
\end{equation}

and the corresponding error:

\begin{equation}
\delta N_{ij}=\sqrt{{1\over N_C}+{1\over N_I}-{1\over N_R}}
\end{equation}

Following \cite{Colless1989} the composite LF can be calculated by:

\begin{equation}
N_{cj}={N_{c0} \over m_j} \sum_i {N_{ij}\over N_{i0}}
\end{equation}

where $N_{cj}$ is the number of cluster galaxies in magnitude bin $j$, and the sum is carried
over the $i$ clusters and $m_j$ is the number of clusters contributing to magnitude bin $j$. Here
$N_{i0}$ is a normalisation factor, corresponding to the (completeness corrected) number of galaxies 
brighter than a given magnitude (here we use $K_s=13$) in each cluster and

\begin{equation}
N_{c0}=\sum_i N_{i0}
\end{equation}

The error is then given by:

\begin{equation}
\delta N_{cj}={N_{c0} \over m_j} \Bigg[ \sum_i \Bigg({\delta N_{ij} \over N_{ij}}\Bigg)^2\Bigg]^{1/2}
\end{equation}

Note that this assumes that the redshift surveys do not select specifically for or against cluster 
members.

\section{Results}

\begin{figure*}
\includegraphics[width=\textwidth]{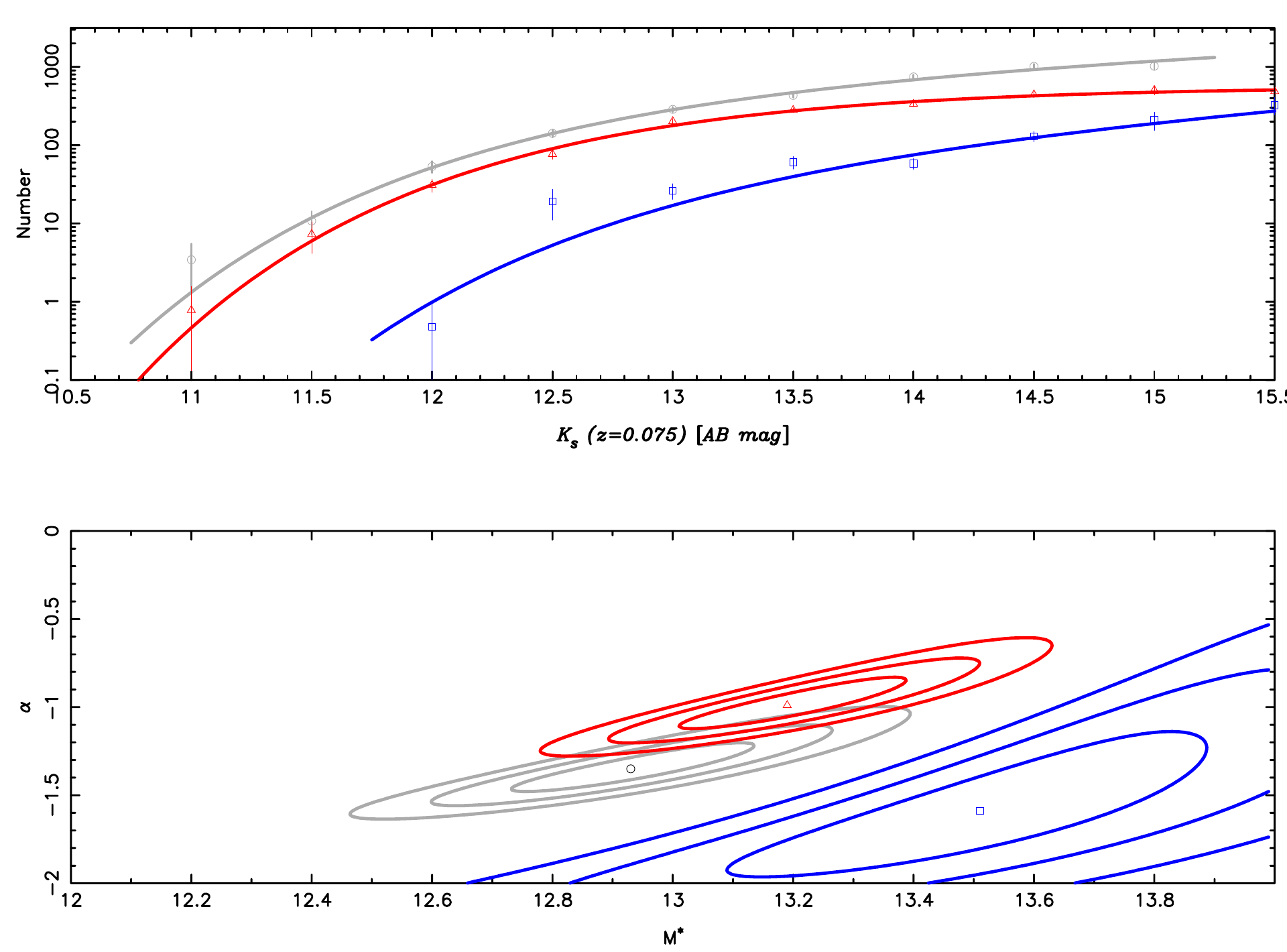}
\caption{The luminosity functions and best fits to the data for galaxies in all 24 clusters (top panel) with
the associated error ellipses (bottom panel). In grey, the total LF, in red the LF for galaxies on the red
sequence and in blue the LF for galaxies in the blue cloud. Tables 2 and 3 show the parameter values.}
\label{lumfs}
\end{figure*}

The best fitting composite $K$-band LF for galaxies in all 24 clusters is shown in Fig.~\ref{lumfs}, assuming a
single Schechter form. The best fitting values are $K^*=12.79\ (M_K=-24.81)\ \pm 0.14$ and 
$\alpha=-1.41 \pm 0.10$. We also show the associated error ellipse as the errors are correlated.

\begin{figure*}
  \includegraphics[width=\textwidth]{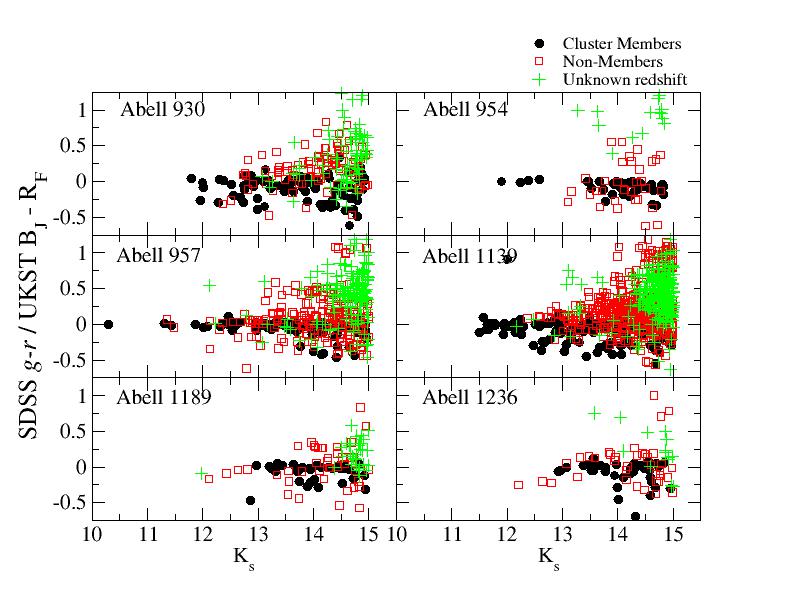}
  \caption{Colour-magnitude diagram for all galaxies in a subsample of clusters (members, non-members and
unknown as in the legend). The colours have already been corrected so that the red sequence has 0 colour.
Some galaxies lie beyond the axis limits in $y$. See Appendix (online) for all other clusters in the sample}
 \label{cmrs}
\end{figure*}

\begin{figure*}
  \includegraphics[width=\textwidth]{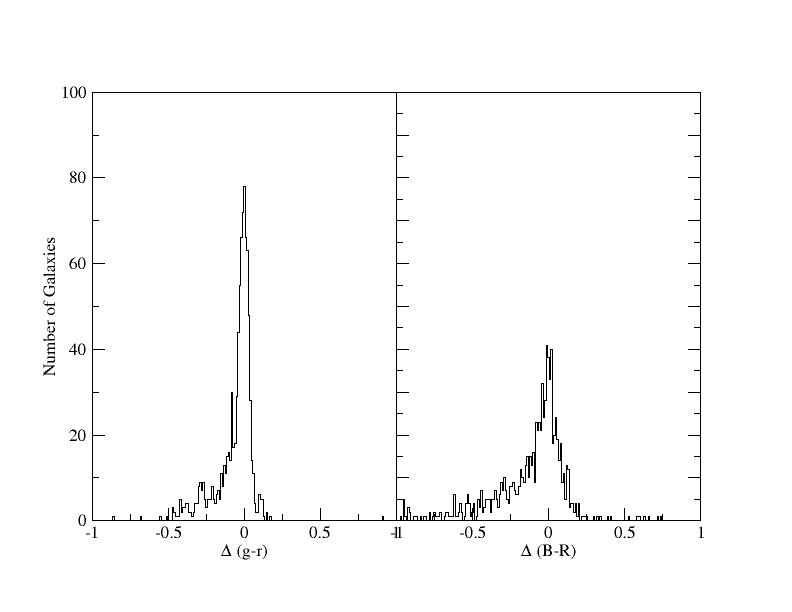}
  \caption{Colour distribution for galaxies about the red sequence, for all 11 clusters with SDSS data (in $g-r$)
and 13 clusters with UKST data (in $B_J-R_F$)}
 \label{dist}
\end{figure*}

Fig.~\ref{cmrs} shows the colour-magnitude relation for galaxies in our clusters, where we have already 
removed the slope and intercept of the red sequence. Colours come from the SDSS or the UKST as 
indicated in Table 1 for each cluster. The slope and intercept of the colour-magnitude relation were 
determined by fitting a minimum absolute deviation straight line to the colour-magnitude relation of 
cluster members, as this minimizes the effects of interlopers on the fit \citep{Beers1990}. The distribution 
of colours about the red sequence for all members is shown in Fig ~\ref{dist}. (panel (a) for clusters with SDSS 
data and (b) for clusters with UKST data). The spread at half maximum on the red edge of the distribution 
is 0.05 mag. for galaxies in clusters with SDSS data ($g-r$) and 0.07 mag. for galaxies in clusters with 
UKST data ($B_J-R_F$). We therefore choose to treat galaxies within $\pm 0.15$ and $\pm 0.21$ mag. 
of the red sequence as red sequence galaxies and the remainder as blue cluster galaxies. Note that
in Fig.~\ref{cmrs} we have not plotted galaxies redder than the red sequence (as defined above) for clarity. 

We then derive a composite LF for red and blue galaxies in all 24 clusters. The red sequence LF and best fit 
are shown in Fig. 2. This has $K^*=13.16$ (--24.44) and $\alpha=-1.00$ (see Table~\ref{par2}). For blue 
galaxies the Schechter function is a poor fit. The best parameters are $K^*=13.51 \pm 0.41$ and 
$\alpha=-1.60 \pm 0.29$, albeit with very large errors (see Fig.~\ref{lumfs})

We can also split our sample according to physically significant cluster properties. The velocity dispersion
may be taken as an indicator of cluster mass. As in our previous paper \citep{DePropris2003} we adopt 
$\sigma=800$ km $s^{-1}$ to separate massive and less massive clusters. We also create samples with
Bautz-Morgan type $>II$ and $\leq II$, where the Bautz-Morgan type reflects the relative dominance of
the brightest galaxies over the rest of the cluster members and may be an indicator of the degree of
dynamical evolution (if the brightest cluster galaxies, for example, grow by dynamical friction and
cannibalism).  Figure~\ref{bmtypes} shows the derived LFs and relative error ellipses. Table~\ref{par1}
below shows the values of the derived parameters.

\begin{table}
\caption{$K$ LF parameters}
\centering
\begin{tabular}{ccc}
\hline\hline
Sample & $K^*$ & $\alpha$  \\
\hline
All & $12.79 \pm 0.14$ & $-1.41 \pm 0.10$ \\
$\sigma < 800$ km s$^{-1}$ & $12.83 \pm 0.25$ & $-1.64 \pm 0.13$ \\
$\sigma > 800$ km s$^{-1}$ & $12.85 \pm 0.18$ & $-1.12 \pm 0.16$\\
BM $\leq$ II & $12.79 \pm 0.32$ & $-1.62 \pm 0.19$ \\
BM $>II$ & $13.10 \pm 0.17$ & $-1.40 \pm 0.14$ \\
\hline\\
\end{tabular}
\label{par1}
\end{table}

\begin{figure*}
  \includegraphics[width=0.93\textwidth]{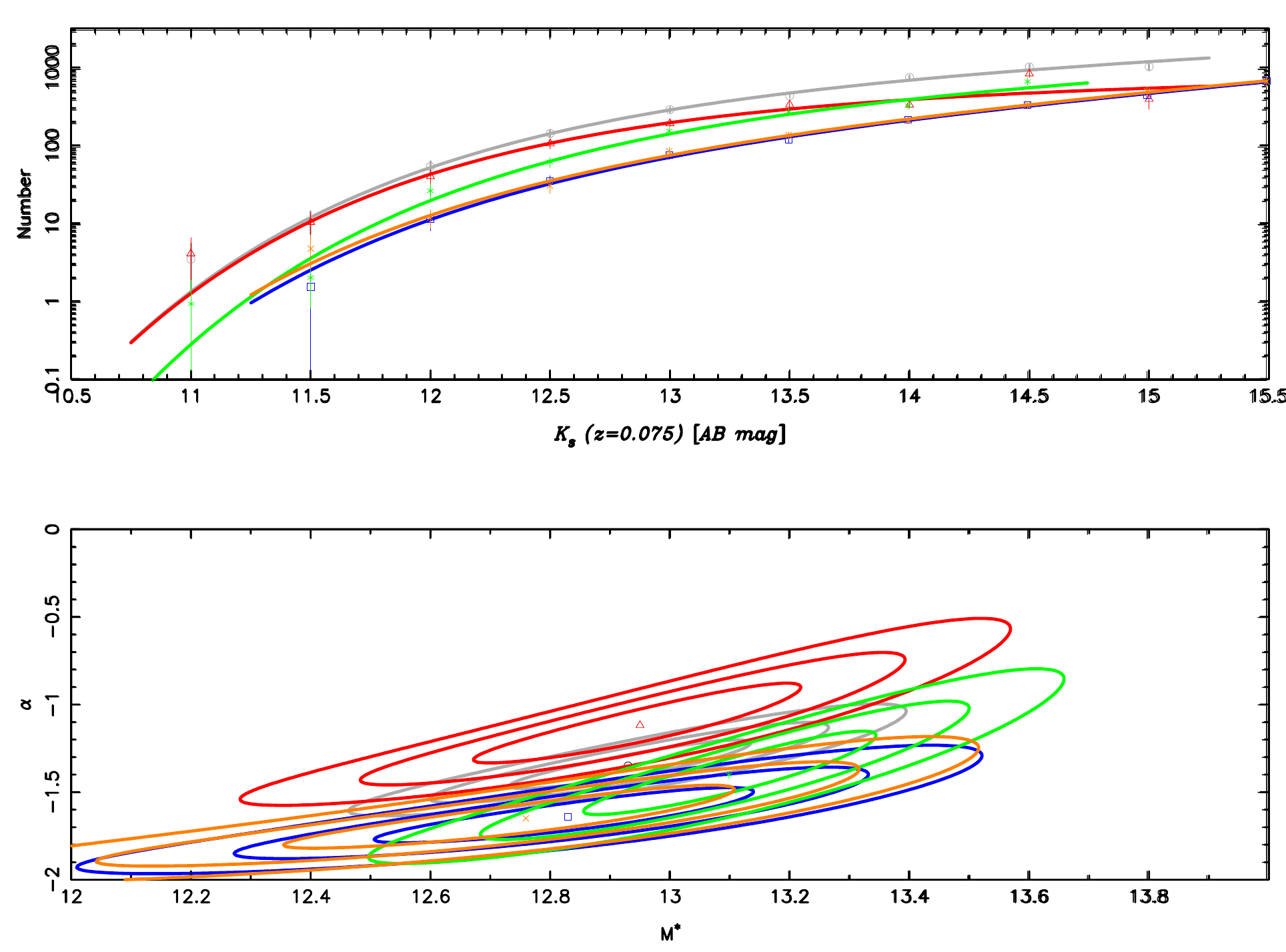}
  \caption{Composite LFs and error ellipses for galaxies in clusters with $\sigma > 800$ km s$^{-1}$ (blue). 
$\sigma < 800$ km s$^{-1}$ (red), BM type $> II$ (green) and BM type $<II$ (orange). The grey line 
is the total LF for all galaxies.}
 \label{bmtypes}
\end{figure*}

For red sequence galaxies, we can also consider subsamples of objects, as for the full sample above. 
This is not possible for the blue cluster members, as there are very few objects  in the sample 
(see Fig.~\ref{lumfs}). The LFs are shown in Fig.~\ref{rslfs} and the relative parameters given in 
Table~\ref{par2}.

\begin{table}
\caption{Red Sequence $K$ LF parameters}
\centering
\begin{tabular}{ccc}
\hline\hline
Sample & $K^*$ & $\alpha$  \\
\hline
All & $13.16 \pm 0.15$ & $-1.00 \pm 0.12$ \\
$\sigma < 800$ km s$^{-1}$ & $13.30 \pm 0.18$ & $-0.99 \pm 0.14$\\
$\sigma > 800$ km s$^{-1}$ & $13.07 \pm 0.24$ & $-1.07 \pm 0.19$\\
BM $\leq$ II & $13.16 \pm 0.21$ & $-1.06 \pm 0.17$ \\
BM $>II$ & $13.38 \pm 0.16$ & $-0.81 \pm 0.16$\\
\hline\\
\end{tabular}
\label{par2}
\end{table}

\begin{figure*}
  \includegraphics[width=\textwidth]{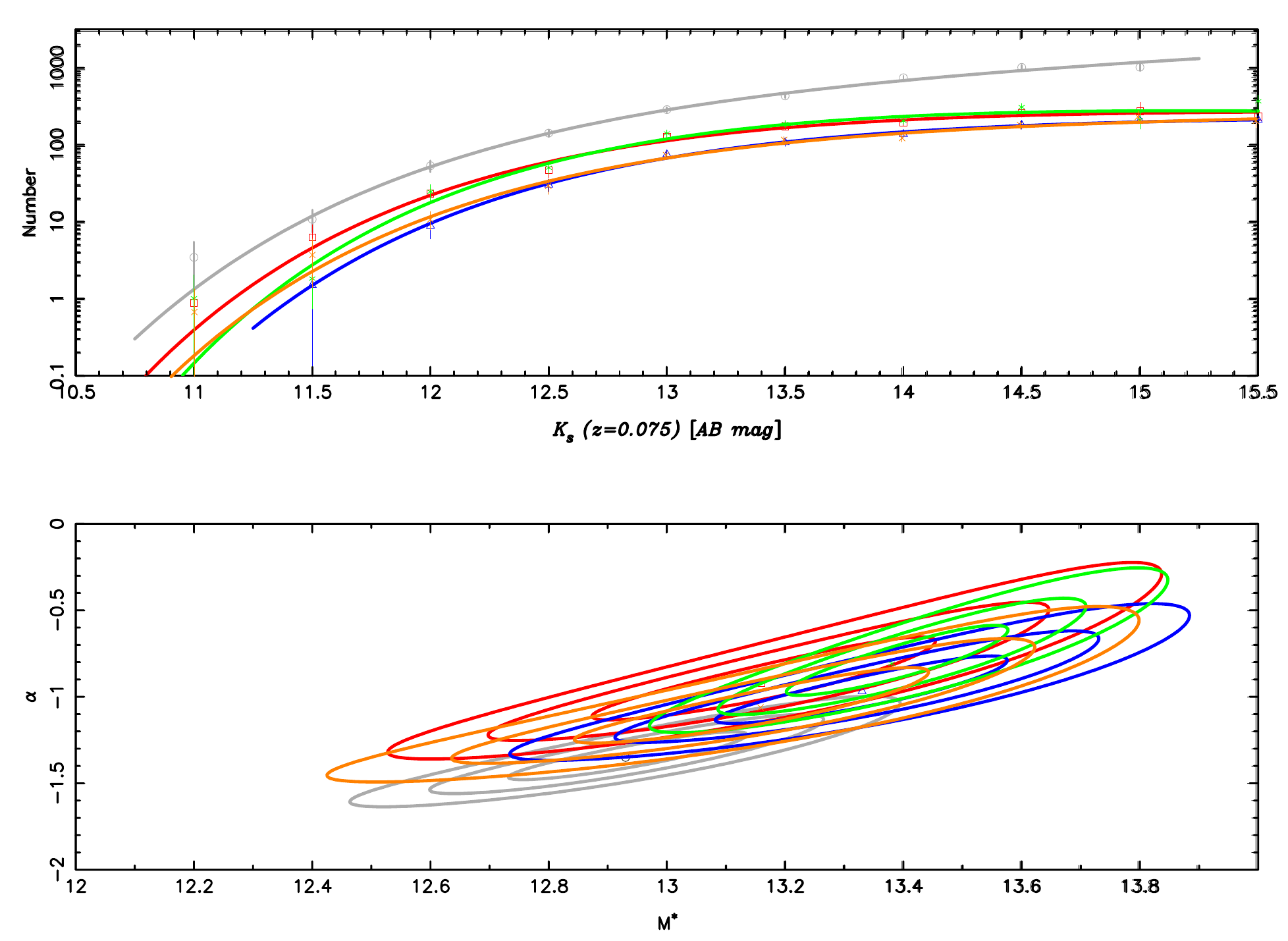}
  \caption{Composite LFs and error ellipses for red sequence galaxies in clusters 
with $\sigma > 800$ km s$^{-1}$ (blue), $\sigma < 800$ km s$^{-1}$ (red), 
BM type $> II$ (green) and BM type $<II$ (orange). The grey line is the total LF.}
 \label{rslfs}
\end{figure*}

\section{Discussion}

\subsection{Comparison with previous work} There are relatively few studies of the $K$-band LF and
most of these have been limited to single or small samples of clusters, owing to the small areas of infrared
detectors until recently. In the following, we have converted all previous results to the concordance cosmology
and to our fiducial redshift $z=0.075$. Our work in Coma \citep{DePropris1998} may be the closest to the
approach we have carried out here, being based on a spectroscopically complete sample of galaxies with 
$H < 14.5$ in the inner $25'$ of the Coma cluster. Assuming $H-K=0.2$ mag. for a $\sim K^*$ galaxy
\citep{Eisenhardt2007}, $K^*_{Coma}=13.57$ in \cite{DePropris1998}, compared to our value of K$^*=12.79$
in the present study. This is within the $2\sigma$ error ellipse shown in Fig.~\ref{lumfs} and closer to the
value we determine for red sequence galaxies (that dominate the core of the Coma cluster). \cite{Skelton2009}
derive $K^*=12.18$, but with a very large error (0.8 mag.) for the Norma cluster. This is still consistent with
our measurement. In our previous analysis of 10 2dF clusters \citep{DePropris2009} we obtained $K^*=12.39$ 
but we have considerably improved our sample, and augmented the redshift completeness, especially at the
faint end. \cite{Merluzzi2010}, for an ensemble of clusters within the Shapley supercluster, has $K^*=12.61$,
closer to our value for 24 clusters here. There is considerable variation when comparing with the $K^*$ values 
for single clusters, as the small number statistics at the bright end makes fitting $K^*$ difficult. However, our
composite LF should provide a better estimate of the LF parameters, as the sparse bright end is more populated. It is not unfortunately not very informative to  carry out a comparison between the individual 
cluster LFs, even with our high completeness, as the errors are so large that there is no statistical power 
in the analysis.

The faint end slope slope we derive is in good to reasonable agreement with previous studies, especially
the estimates by \cite{Skelton2009} for Norma and the Shapley supercluster in \cite{Merluzzi2010}. In Coma, 
the slope is affected by the presence of an inflection in the LF at intermediate magnitudes, but we may remark
that there is good agreement with the slope of the LF of red sequence galaxies (Table~\ref{par1}) - nearly all
spectroscopically confirmed Coma galaxies in \cite{DePropris1998} are red sequence members. The slope 
we derive is also steeper than in our previous analysis for a subset of these clusters \citep{DePropris2009},
although the quality of the data has improved, especially at the faint end, which might provide at least part 
of the explanation for the discrepancy. 

\subsection{Comparison with the general field}

The most recent field $K$-band LF by \cite{Jones2006} has $K^*=13.06$ (for our cosmology and  at
$z=0.075$) and $\alpha=-1.16$, although the fit to a Schechter function is not good. \cite{Bell2003}
has $K^*=13.60$ and $\alpha=-0.77$, while \cite{Cole2001} has $K^*=13.45$  and $\alpha=-0.96$. 
Despite the differences between these studies, it appears that the cluster $K$ band LF is slightly brighter 
(by about 30\%) and somewhat steeper than in the general field. The brighter LF for cluster galaxies when
compared to the field was also found in \cite{DePropris2003}, among others, and is understandable if the 
brighter cluster members are formed from mergers within the cluster environment and/or if the richer cluster
environment favours the formation of more massive systems. In order to understand which mechanism is 
dominant, or even if several such processes are operating, one would need to compare field and cluster 
LFs as a function of redshift. One caveat to this conclusion is that most previous work has been based
on 2MASS photometry. As we have seen above, and as previously found by \cite{Andreon2002} and \cite{
Kirby2008}, 2MASS magnitudes seem to be $\sim 0.3$ magnitudes fainter, likely because of light losses
in the low surface brightness envelopes of galaxies. With this, the difference between field and cluster
$K^*$ is much reduced and they appear to be identical within the errors. More accurate photometry for
wide field surveys is needed to resolve this issue.

The existence of steeper LFs in clusters vs. field galaxies were already pointed out by \cite{Merluzzi2010} for 
the Shapley supercluster. This steeper slope is somewhat surprising. This may be an environmental effect,
although we would expect the field LF to be steeper if the trends we observe in clusters (see below) continue.
One possibility is that the 2MASS photometry used by \cite{Cole2001,Bell2003} and \cite{Jones2006} 
systematically misses faint galaxies, especially at lower surface brightness levels, and thus leads to a flatter 
slope than would otherwise be measured (e.g., \citealt{Andreon2002,Kirby2008}). Otherwise, one would have 
to find a  mechanism by which dwarfs are preferentially formed or preserved in the theoretically more hostile 
cluster environments.

\subsection{Environmental Effects}

We consider the effects of the cluster environment by splitting our sample into several subsamples
according to physically significant properties of clusters such as velocity dispersion and the Bautz-Morgan
class. We do not find strong evidence that $K^*$ varies across these subsamples. This argues that
the environment does not strongly affect the behaviour of bright galaxies, at least within clusters.

The only significant difference we find is for the slope of the LF of $\sigma < 800$ km s$^{-1}$ subsample
to be steeper than for the $\sigma > 800$ km s$^{-1}$ subset, at nearly the $3\sigma$ level (as shown in
Fig.~\ref{bmtypes}. For clusters having  $\sigma > 800$ km s$^{-1}$ the LF is flatter than the total LF,
while for cluster with $\sigma < 800$ km s$^{-1}$ it is significantly steeper. We also find that the LF for
clusters of BM type $<$ II has also a significantly steeper slope than the total LF and is different from the
LF for clusters with BM type $>$ II. The two LFs (for clusters with $<800$ km s$^{-1}$ and for clusters
with BM type $<II$) are also very similar to each other. Although the two samples do not fully overlap.
clusters with BM type $>$ II have slightly larger velocity dispersions than clusters with BM type $\leq$ II.

This may suggest that relatively low mass systems, possibly with a single dominant galaxy, are more 
favourable to the formation or survival of dwarf galaxies (as in the low density environments of \citealt{
Merluzzi2010}). One possibility is that as clusters grow (from accretion of single galaxies and groups, or mergers with other clusters) the relative dominance of brightest cluster galaxies decreases as more 
luminous systems are included within the cluster, and as this process takes place dwarfs may be 
destroyed or captured by more massive galaxies. Here, a single or two massive dominant galaxies 
in the cluster may be a sign of dynamical youth, rather than evolution. As clusters grow, mergers 
may become rarer (as the velocity dispersion increases) and the brighter end of the LF may be `filled 
in' by the infall of bright galaxies. 

We have also derived the LF for red sequence galaxies, both for all clusters and for the subsamples 
we have defined above. The red sequence LF is flatter than the total LF. The red sequence LF does 
not appear to vary significantly between any of the cluster subsamples. This suggests that any 
environmental variation is due to the different contributions to the faint end of the LF from blue 
galaxies. These dominate the low mass end of the LF. We are not able to explore the environmental
dependence of the blue LF as the statistics are too poor (the normalisation of the blue LF is about 
one order of magnitude lower than the red LF). However, the LF slope is similar to that of clusters
with $\sigma < 800$ km s$^{-1}$ and BM type $<$ II. This would suggest that these star-forming 
dwarf galaxies are preferentially preserved in low mass environments or are destroyed in higher
density systems, with many large giants, rather than a single dominant system -- it is possible that 
most of the growth of brightest cluster galaxies may take place outside of clusters - an intriguing
parallel may be offered by the `infalling' cD galaxy in the Coma NGC4839 subgroup or the J0454-0309
fossil subgroup \citep{Schirmer2010}.

The actual evolution of the blue galaxies is interesting to consider. Because the red sequence LF
is much flatter, they cannot be easily added on to the red sequence (e.g., by quenching). They might
fade considerably, and contribute to the steeper faint end observed by (e.g.) \cite{Moretti2015}, but
one does not expect much fading in $K$, unless they also lose considerable mass (e.g., to tidal
stripping). It would be interesting to increase the sample of clusters and obtain deeper spectroscopy.

\cite{Merluzzi2010} found evidence for an environmental dependence on the near-infrared LF in
the Shapley supercluster, with the slope of the LF increasing towards lower density regions and
being steeper than in the field. We did not find strong evidence that the LF varies radially in our
earlier work \citep{DePropris2003} but the statistics at large ($> 300$ kpc) radii were quite small.
\cite{Adami2007} suggests that the LF of the Coma cluster steepens along the North-South axis
corresponding roughly to the infall direction, while \cite{Boue2008} claim that the LF in Abell 496
steepens towards its outer regions. In Abell 119 \cite{Lee2016} observe a steepening of the LF 
towards the outer low density regions, together with a more pronounced dip at intermediate
luminosities. The trend of the LF to become steeper in bluer bands is well known from several
studies (see for instance \citealt{mcnaught2015}). This would suggest that most of the ``environmental'' 
variation originates from the quenching of low mass galaxies in relatively low density regions, whereas 
the red sequence members have been largely preprocessed before the epoch of observation (e.g.,
\citealt{Gilbank2008, Zirm2008}). This would explain the observation that the red sequence does 
not very between our cluster subsamples. Red sequence galaxies are already processed into 
cluster members in lower density environments.

We can finally compare our findings with the models in \cite{Vulcani2014}. For increasing halo mass
we see that the $L^*$ stellar mass increases by about 30\% over two orders of magnitude in halo mass;
this is broadly consistent with the field vs. cluster comparison above, but not with the nearly constant
$K^*$ in all clusters we consider. However, the local field values may be affected by problems with 2MASS
photometry. Similarly, the slope is very well matched by the models, but the change in slope with environment
is not. As pointed out by \cite{Vulcani2014} the models may still suffer from several shortcomings, especially
in the inclusion of cluster-specific environmental effects, the efficiency of galaxy formation and the evolution
of central and satellite galaxies.

\section*{Acknowledgements}

Funding for SDSS-III has been provided by the Alfred P. Sloan Foundation, the Participating Institutions, the National Science Foundation, and the U.S. Department of Energy Office of Science. The SDSS-III web site is http://www.sdss3.org/

This research has made use of data obtained from the SuperCOSMOS Science Archive, prepared and hosted 
by the Wide Field Astronomy Unit, Institute for Astronomy, University of Edinburgh, which is funded by the 
UK Science and Technology Facilities Council.

This research has also made use of the UKIRT Infrared Deep Sky Survey data base.

This research has made use of the NASA/IPAC Extragalactic Database (NED) which is operated by the Jet Propulsion Laboratory, California Institute of Technology, under contract with the National Aeronautics and Space Administration.

We thank the anonymous referee for a very comprehensive report that has helped improve this article.











\appendix
\section{Appendix}
\begin{figure*}
  \includegraphics[width=\textwidth]{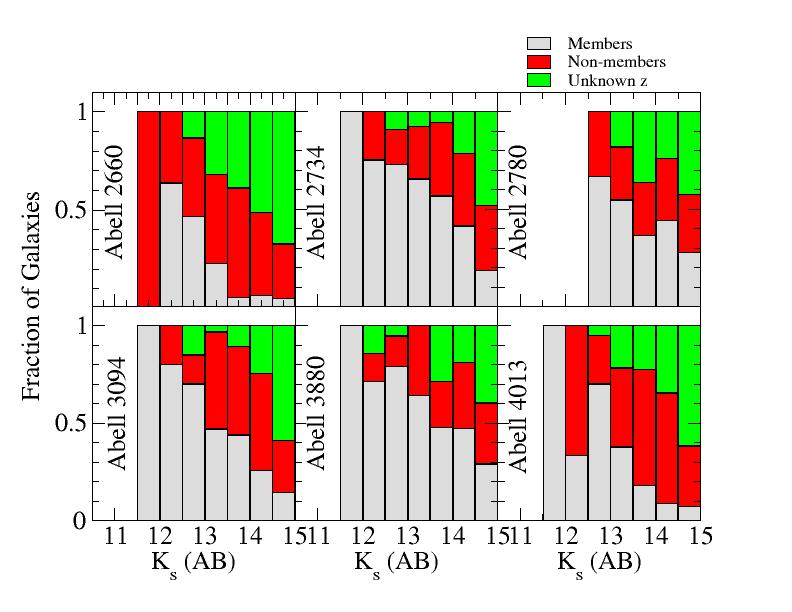}
 \caption{Continued.}
\setcounter{figure}{0}
\end{figure*}

\begin{figure*}
  \includegraphics[width=\textwidth]{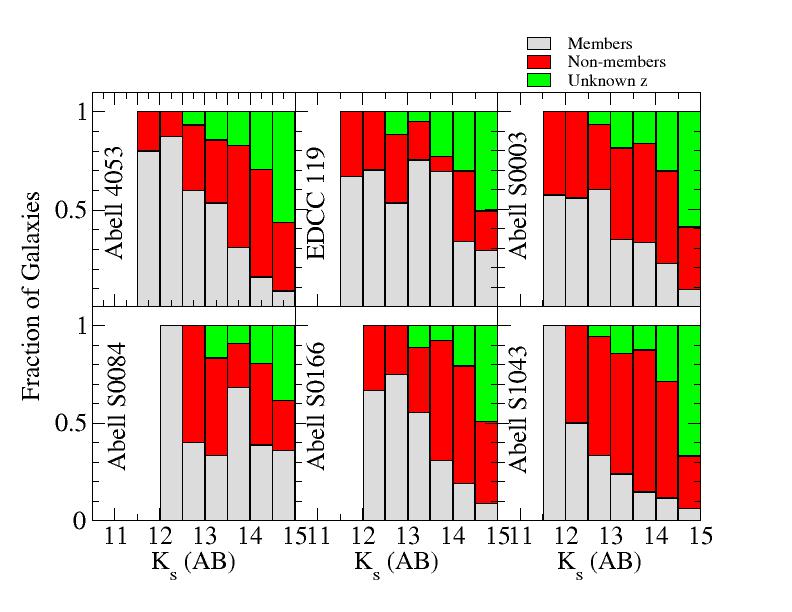}
 \caption{Continued.}
\setcounter{figure}{1}
\end{figure*}

\begin{figure*}
  \includegraphics[width=\textwidth]{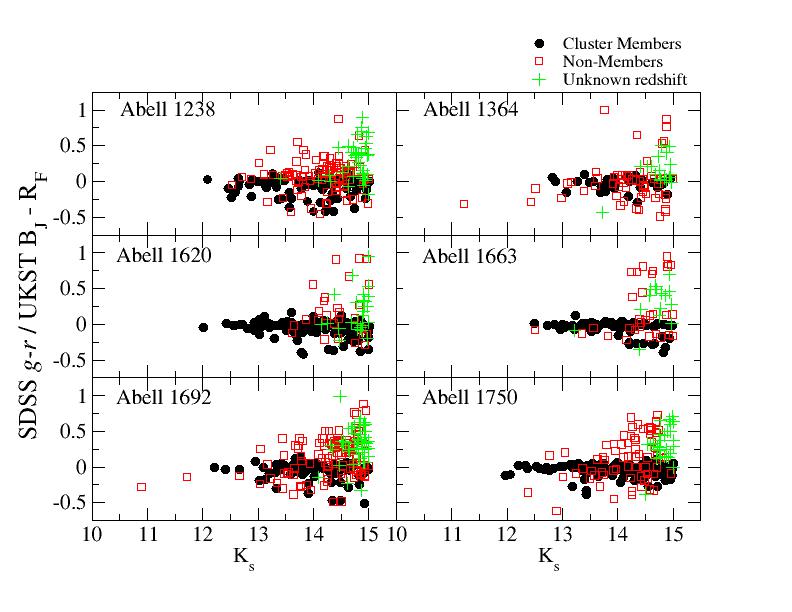}
  \caption{Colour-magnitude diagram for a subset of clusters (members, non-members and
unknown redshifts as in the legend). The colours have already been corrected so that the red sequence 
has 0 colour. See Appendix for all other clusters in the sample}
\setcounter{figure}{1}
\end{figure*}

\begin{figure*}
  \includegraphics[width=\textwidth]{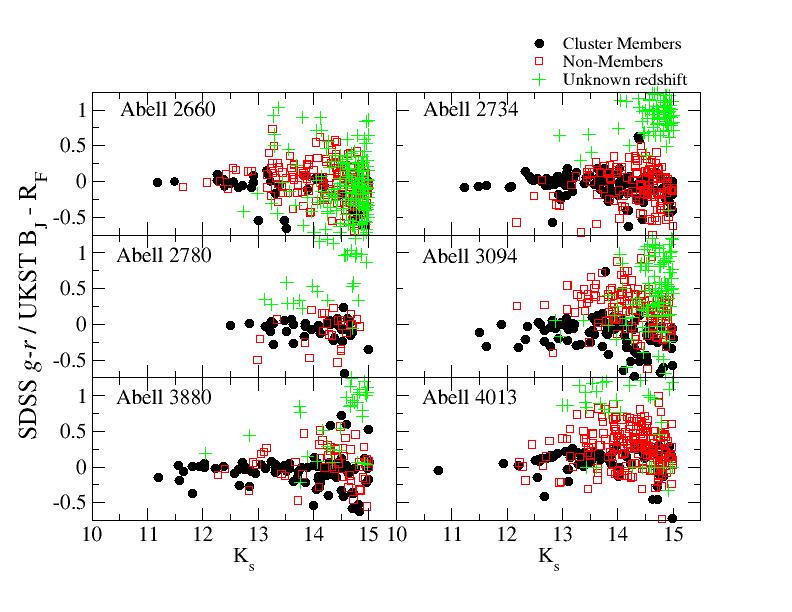}
  \caption{Continued}
\setcounter{figure}{1}
\end{figure*}

\begin{figure*}
  \includegraphics[width=\textwidth]{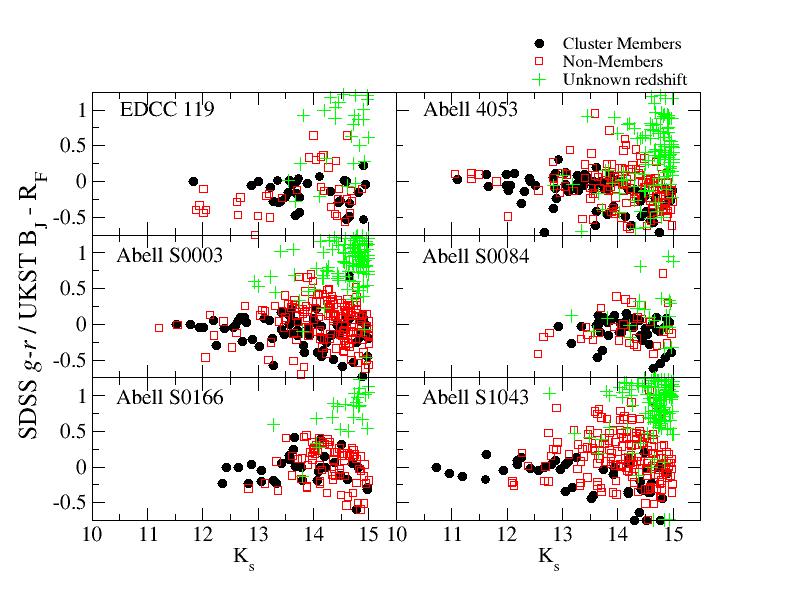}
  \caption{Continued}
\setcounter{figure}{1}
\end{figure*}

\bsp	
\label{lastpage}
\end{document}